\documentstyle[prd,aps,multicol,psfig]{revtex}

\renewcommand{\narrowtext}{\begin{multicols}{2} \global\columnwidth20.5pc}
\def\al{\alpha}
\def\be{\beta}
\def\ga{\gamma}
\def\de{\delta}

\def\ve{\varepsilon}

\def\et{\eta}

\def\la{\lambda}

\def\rh{\rho}

\def\si{\sigma}

\def\ta{\tau}

\def\fr#1#2{{{#1} \over {#2}}}
\def\half{{\textstyle{1\over 2}}}

\def\frac#1#2{{\textstyle{{#1}\over {#2}}}}

\def\lsim{\mathrel{\rlap{\lower4pt\hbox{\hskip1pt$\sim$}}
    \raise1pt\hbox{$<$}}}
\def\gsim{\mathrel{\rlap{\lower4pt\hbox{\hskip1pt$\sim$}}
    \raise1pt\hbox{$>$}}}
\def\sqr#1#2{{\vcenter{\vbox{\hrule height.#2pt
         \hbox{\vrule width.#2pt height#1pt \kern#1pt
         \vrule width.#2pt}
         \hrule height.#2pt}}}}

\def\prt{\partial}
\def\lrpartial{\raise 1pt\hbox{$\stackrel\leftrightarrow\partial$}}

\def\etal{{\it et al.}}

\newcommand{\beq}{\begin{equation}}
\newcommand{\eeq}{\end{equation}}
\newcommand{\bea}{\begin{eqnarray}}
\newcommand{\eea}{\end{eqnarray}}
\newcommand{\rf}[1]{(\ref{#1})}

\begin{document}

\title{Threshold analyses and Lorentz violation} 

\author{
Ralf Lehnert}

\address{CENTRA, Department of Physics, Universidade do Algarve,
8000-117 Faro, Portugal}

\maketitle

\begin{abstract}
In the context of threshold investigations of Lorentz violation, 
we discuss the fundamental principle of coordinate invariance, 
the role of an effective dynamical framework, 
and the conditions of positivity and causality. 
Our analysis excludes a variety of previously considered 
Lorentz-breaking parameters 
and opens an avenue for viable dispersion-relation investigations 
of Lorentz violation. 
\end{abstract}

\narrowtext

\section{Introduction}

For several decades, 
a sizeable amount of theoretical work 
has been directed 
towards a unified quantum description 
of all fundamental interactions including gravity. 
Comparable experimental efforts 
have been inhibited, 
partly by the expected minuscule size of Planck-scale effects, 
and partly by the absence of a realistic underlying framework. 
One approach to overcome this phenomenological obstacle is 
to identify exact symmetries in present-day physics 
that may be violated in underlying theories 
and are amenable to high-precision tests. 

Lorentz and CPT invariance satisfy these criteria. 
They are cornerstones of our present understanding of nature 
at the fundamental level, 
and certain Lorentz and CPT tests 
belong to the most precise null experiments available. 
Moreover, 
a mechanism that could cause spontaneous Lorentz and CPT breaking 
in the framework of string field theory 
was discovered 
more than a decade ago \cite{kps}. 
Subsequent investigations 
have considered the possibility of Lorentz violation 
in other contexts, 
such as spacetime foam \cite{ell98,suv}, 
nontrivial spacetime topology \cite{klink}, 
loop quantum gravity \cite{amu}, 
realistic noncommutative field theories \cite{chklo}, 
and spacetime-varying couplings \cite{klp03}. 

Low-energy effects 
of Lorentz breaking 
are described 
by the Standard-Model Extension (SME) 
\cite{ck97,ck98,jk,kl01,klp02}. 
This dynamical framework 
is constructed to contain all Lorentz- and CPT-violating 
lagrangian terms consistent with coordinate invariance, 
a fundamental requirement to be discussed below. 
The SME 
has provided the basis 
for numerous experimental Lorentz- and CPT-violation searches 
involving hadrons 
\cite{hadronexpt,hadronth,baryo}, 
protons and neutrons 
\cite{pn,pnhh,pnthhh}, 
electrons 
\cite{pnthhh,eexpttrap,ethtrap,ethpol,eexptpol}, 
photons 
\cite{cfj,km,km02,brax,lipa}, 
muons \cite{muons}, 
and neutrinos \cite{ck97,cg99,neutrinos}. 

Within the SME, 
it is straightforward to confirm 
that spacetime-symmetry breakdown 
generally modifies one-particle dispersion relations 
\cite{cfj,ck97,ck98,kl01}. 
This feature of Lorentz violation 
permits the prediction of possible experimental signatures 
based purely on kinematical arguments. 
For example, 
primary ultra-high-energy cosmic rays (UHECR) 
at energies eight orders of magnitude 
below the Planck mass have been observed. 
At these scales, 
Lorentz-breaking effects 
might be amplified 
relative to the ones in low-energy experiments 
leading to potentially observable 
threshold modifications for particle reactions. 
This idea 
has been adopted in a number of recent investigations 
\cite{ell98,cg99,klu,kif,bert,alo,gac,jlm,koma,vankov,alf,frey,ghar,posp}. 
In many of these studies, 
however, 
the dispersion relations are constructed phenomenologically 
without reference to an effective dynamical framework 
and other physical principles.  

The goal of the present work 
is to investigate 
how some of the arbitrariness 
in the construction of Lorentz-violating dispersion relations 
can be removed. 
Our investigation 
is based on the principle of coordinate invariance 
and on the condition of compatibility 
with an effective dynamical framework. 
It is argued 
that these two features are fundamental enough 
for being physical requirements, 
while maintaining relative independence 
of the details of the Planck-scale theory. 
Moreover, 
the implementation of general dynamical properties 
significantly increases 
the scope of threshold investigations.  
We also discuss positivity and causality, 
properties which further add 
to the viability of kinematical analyses. 
Throughout we assume energy-momentum conservation. 

In Sec.\ 2, 
we comment on the requirement of coordinate independence 
and its consequences for dispersion relations. 
Section 3 discusses 
dispersion relations 
from the viewpoint of compatibility 
with the SME. 
In Sec.\ 4, 
issues regarding positivity and causality are addressed. 
Further results and a discussion of sample dispersion relations 
often considered in the literature 
can be found in Sec.\ 5. 
A brief summary is contained in Sec.\ 6. 

\section{Coordinate independence} 

Coordinate independence 
is essential in physics, 
and its role in the context of Lorentz violation 
is well established \cite{ck97,ck98}. 
However, 
many threshold analyses lack coordinate invariance, 
and occasionally Lorentz breaking is identified 
with the loss of coordinate independence. 
It is therefore appropriate to review some arguments 
behind this fundamental principle 
before discussing its consequences for dispersion relations. 

A certain labeling scheme for events in space and time 
corresponds to a choice of coordinate system. 
Such a labeling of events 
is a pure product of human thought 
and thus arbitrary to a large extent. 
Despite being one of the most common and important {\it tools} 
in physics, 
coordinate systems fail to possess {\it physical reality} 
in the sense that the physics must remain independent 
of the choice of coordinates. 
This principle, 
also called observer invariance, 
is one of the most fundamental in science. 
Mathematically, 
observer invariance can be implemented 
by choosing a spacetime manifold 
as the arena for physical events 
and certain tensors or spinors 
for the representation of physical quantities. 

We also mention 
that coordinate invariance is much more general 
than Lorentz symmetry. 
For example, 
nonrelativistic classical mechanics and Newton's law of gravitation 
fail to be Lorentz invariant 
but can be formulated in the coordinate-free language of three-vectors. 
Only when the spacetime manifold is taken to be lorentzian, 
the Lorentz transformations acquire a significant role: 
They implement changes between local Minkowski frames. 

Lorentz violation 
is associated with nontrivial vacua 
described at low energies by nondynamical tensorial backgrounds. 
A tensor background 
can lead, 
for instance, 
to direction-dependent particle propagation, 
a situation comparable to the one inside certain crystals. 
One then says 
that particle Lorentz invariance is broken \cite{ck97}. 
However, 
observer Lorentz invariance 
remains fully intact, 
so that 
locally one can still work 
with the metric $\et^{\mu\nu}={\rm diag}(1,-1,-1,-1)$, 
particle four-momenta 
still transform in the usual way under coordinate changes, 
and the conventional tensors and spinors 
still represent physical quantities. 

The important difference 
between observer and particle Lorentz symmetry 
can be illustrated 
in the conventional context 
of a classical point particle of mass $m$ and charge $q$ 
in an external electromagnetic field $F^{\mu\nu}$ obeying
\beq
m\:\fr{dv^{\mu}}{d\ta}=qF^{\mu\nu}v_{\nu}\; .
\label{example}
\eeq
Here, $v^{\mu}$ is the four-velocity of the particle 
and $\ta$ is its proper time. 
In general, 
invariance under rotations of the particle's trajectory 
(and thus particle Lorentz symmetry) 
is broken by the external $F^{\mu\nu}$ 
resulting in the nonconservation of the charge's angular momentum, 
for instance. 
However, 
Eq.\ \rf{example} is a tensor equation 
valid in all coordinate systems 
maintaining observer Lorentz symmetry. 
We remark that in the above example, 
the background $F^{\mu\nu}$ is a local electromagnetic field 
generated by other four-currents 
that can in principle be controlled. 
On the other hand, 
a background in the Lorentz-violating context 
is a global property of the effective vacuum 
outside of experimental control. 

{\bf Observer-invariant dispersion-relation ansatz.}
In the published literature, 
Lorentz-violating dispersion relations 
are usually taken to be of the form \cite{comment}
\beq
{\la_0}^2-{\vec{\la}}^{\:2}=m^2+\de f(\la_0,\vec{\la})\; ,
\label{lvdr}
\eeq
where $m$ is the usual mass parameter 
and $\la^{\mu}=(\la^0,\vec{\la})$ 
is the plane-wave four-vector 
(before the reinterpretation of negative energies, 
see Eq.\ \rf{reint}). 
The function $\de f(\la_0,\vec{\la})$ 
controls the extent of the Lorentz-breaking. 
We proceed under the assumption 
that the dynamics of the free particle is described 
by a linear partial differential equation 
with constant coefficients. 
In the absence of nonlocalities,  
we obtain the polynomial ansatz 
\beq
\de f(\la_0,\vec{\la})=
\sum_{n\ge 1}\hspace{4mm}
\overbrace{\hspace{-3.5mm}T_{(n)}^{\;\al\be\;\cdots\;\;}}^{n\; \rm indices}
\hspace{-1mm}\underbrace{\hspace{.5mm}\la_{\al}\la_{\be}\;\cdots\;}
_{n\; \rm factors} \;\; .
\label{ansatz}
\eeq
Here, $T_{(n)}^{\;\al\be\;\cdots\;}$ denotes a constant tensor of rank $n$ 
parametrizing the violation of particle Lorentz symmetry. 
All the tensor indices $\al,\be,\,\ldots$ 
are understood to be distinct 
but each one is contracted with a momentum factor, 
so that all terms in the sum are observer Lorentz invariant. 
Note that the $T_{(n)}$ can be taken as totally symmetric. 
The above arguments are summarized in Result (i): 
{\it Lorentz-violating dispersion relations 
must satisfy the the requirement of coordinate independence, 
and therefore they hold in any frame 
and contain only nonnegative integer powers 
of $\la_0$ and $|\vec{\la}|$} \cite{fn3}.

{\bf Energy degeneracy.} 
For a given $\vec{\la}$,  
Eq.\ \rf{lvdr} is a polynomial in $\la_0$, 
which has generally multiple roots 
lifting the conventional energy degeneracy 
between particle, antiparticle, and possible spin-type states. 
This is intuitively reasonable 
because degeneracies usually arise through symmetries, 
and in the present context the number of symmetries 
is normally reduced. 
This raises the question 
what degree of physical Lorentz violation 
is described by dispersion-relation modifications 
preserving the usual degeneracy. 
This issue assumes particular importance 
in light of the fact 
that the majority of threshold analyses 
employ dispersion relations maintaining the conventional degeneracy. 

For example, 
it is known 
that ``doubly special relativities'' \cite{dsr1,dsr2} 
maintain the conventional number of spacetime symmetries 
and exhibit the ordinary equality of all particle and antiparticle energies 
for a given three-momentum. 
However, 
such approaches to the loss of Lorentz symmetry 
appear to be physically indistinguishable 
from  the conventional case \cite{triv}. 
In the present context, 
degeneracy is maintained,
e.g., 
when $\de f(\la_0,\vec{\la})$ 
contains only ${\la_0}$-independent and ${\la_0}^2$ terms. 
In certain models, 
one can then find transformations 
removing the Lorentz breaking 
from the particle species in question 
at the cost of introducing the violations 
in a different sector of the theory \cite{km02}. 
Then, 
Lorentz-violating and Lorentz-symmetric sectors 
can have the same  number of spacetime symmetries 
and associated conserved currents 
but the symmetry generators and the conserved quantities 
differ in the two sectors. 
This could potentially lead 
to additional conservation constraints on particle reactions. 

The above arguments support the conjecture 
that Lorentz violation can be completely removed from a theory 
with identical, degeneracy-preserving dispersion-relation corrections 
in all sectors. 
In summary, 
we are lead to Result (ii):
{\it For generality and to avoid possible triviality, 
degeneracy-lifting dispersion relations 
must be included into kinematical investigations.}
We remark 
that terms with more than four powers of $\la_0$ 
generally result in more than four particle states, 
which could lead to interpretational difficulty. 

{\bf Rotational symmetry.}
Most threshold analyses make the simplifying assumption 
of rotation invariance in certain frames, 
and then $\de f(\la_0,\vec{\la})$ is expanded in powers of $|\vec{\la}|$ 
with $\la_0$ contributions absent. 
We remark in passing that this procedure 
is associated with potential triviality problems 
because the particle energies remain degenerate. 
But there is also another unsatisfactory aspect of this approach. 
Many authors include odd powers of $|\vec{\la}|$ 
into the expansion of $\de f(|\vec{\la}|)$ 
violating coordinate independence: 
The implementation of rotation symmetry requires 
that the $T_{(n)}$ are constructed 
from $\et^{\mu\nu}$ and constant tensors 
with vanishing spacelike and mixed components. 
Then, 
odd powers of $|\vec{\la}|$ are absent 
because the spacelike and mixed contributions 
to $T_{(n)}$ are of $\de^{jk}$-type, 
so that $\vec{\la}$ always appears contracted with itself. 
For example, 
the respective dispersion-relation corrections 
for $n=1,\ldots,4$ (up to constants) are: 
\bea
&& \la_0\; ,\nonumber\\
&& {\la_0}^2\; ,\quad \vec{\la}^{\:2}\; ,\nonumber\\
&& {\la_0}^3\; ,\quad \la_0\vec{\la}^{\:2}\; , \nonumber\\
&& {\la_0}^4\; ,\quad {\la_0}^2\vec{\la}^{\:2}\; ,\quad \vec{\la}^{\:4}\; .
\label{rotcont}
\eea

For completeness, 
we mention 
that certain special choices of the tensors $T_{(n)}$ 
could generate effective corrections 
with odd powers of $|\vec{\la}|$. 
For instance, 
consider a case 
with an {\it unsuppressed} correction $\de f(\la_0,\vec{\la})$ 
leading to the observer-invariant dispersion relation 
$\la^4-2m^2\la^2+m^4=N^2 |\vec{\la}|^{2j}$, 
where $\la^2=\la^{\nu}\la_{\nu}$, $N$ is a parameter for Lorentz breaking, 
and $j$ is an integer. 
Taking the square root yields 
$\la_0^2-{\vec{\la}}^{\:2}=m^2\pm N |\vec{\la}|^{j}$. 
For appropriate values of $j$, 
odd powers of $|\vec{\la}|$ are generated. 
Note, however, 
the partial nondegeneracy. 
This yields Result (iii): 
{\it Odd powers of $|\vec{\la}|$ 
are excluded in dispersion-relations 
modifications $\de f(\la_0,\vec{\la})$ 
when rotational invariance is assumed} \cite{fn6}. 
An explicit example, 
which also demonstrates 
that this result cannot be bypassed 
by the common substitution $\la_0\rightarrow|\vec{\la}|$, 
is discussed in Sec.\ 5. 

\section{The standard-model extension}

Implementing general dynamical features 
significantly increases the scope 
of particle-reaction analyses. 
However, 
the use of dynamics in threshold investigations 
has recently been questioned 
on the grounds of framework dependence \cite{amca}. 
We disagree with this claim  
and begin with a few remarks about purely kinematical analyses. 

Although kinematics imposes powerful constraints on particle reactions, 
it provides only an incomplete description of reaction processes: 
An expected high-energy reaction can be suppressed 
by modified dispersion relations 
but also by novel symmetries, for example. 
Similarly, 
the presence at high energies 
of a reaction kinematically forbidden at low energies 
could perhaps be explained by additional channels due to the loss of 
low-energy symmetries or new undetected particles. 
Moreover, 
models of both acceleration mechanisms for UHECRs 
and shower development in the atmosphere involve 
conventional dynamics. 
Thus, in studying threshold bounds on Lorentz violation, 
assumptions outside kinematics 
such as dynamical quantum-field aspects 
cannot be eliminated completely. 

Possible dynamical features are constrained 
by the requirement 
that known physics must be recovered 
in certain limits, 
despite some freedom 
in introducing dynamics 
compatible with a given set of kinematical rules. 
Moreover, 
it appears difficult 
and may even be impossible 
to find an effective theory containing the Standard Model 
with dynamics significantly different from the SME. 
We also mention 
that kinematics investigations 
are limited to only a few 
potential Lorentz-violating signatures 
from fundamental physics. 
From this viewpoint, 
it is desirable 
to explicitly implement dynamics of sufficient generality 
into the search for Lorentz breaking.  

{\bf The generality of the SME.}
To appreciate 
the generality of the SME, 
we review the philosophy behind its construction \cite{ck97,ck98}. 
One adds to the usual Standard-Model lagrangian ${\cal L}_{\rm SM}$ 
Lorentz-violating terms $\de {\cal L}$: 
\beq
{\cal L}_{\rm SME}={\cal L}_{\rm SM}+\de {\cal L}\; .
\label{sme}
\eeq
Here, ${\cal L}_{\rm SME}$ denotes the lagrangian of the SME. 
The correction $\de {\cal L}$ 
is formed by contracting standard-model field operators 
of unrestricted dimensionality 
with Lorentz-violating tensorial parameters 
yielding observer Lorentz scalars. 
It is thus apparent 
that the complete set of possible contributions to $\de {\cal L}$ 
yields the most general effective dynamical description 
of Lorentz breaking 
at the level of observer Lorentz-invariant quantum field theory. 

Note that 
instead of constructing corrections to the energy-momentum relation, 
one proceeds at the level of the lagrangian. 
This superior approach yields a much more powerful 
and dynamical framework, 
despite employing a philosophy 
that is at its base {\it de facto} identical 
to the dispersion-relation case 
considered in the previous section. 

Potential Planck-scale features, 
such as a certain  discreteness of spacetime 
or a possible non-pointlike nature of elementary particles, 
are unlikely to invalidate 
the above effective-field-theory approach 
at present energies. 
On the contrary, 
the extremely successful Standard Model is normally viewed 
as an effective field theory 
approximating more fundamental physics. 
If the underlying theory 
indeed incorporates minute Lorentz-violating effects, 
it would appear contrived 
to consider low-energy effective models 
outside the framework of quantum fields. 
We mention in passing 
that Lorentz-symmetric aspects 
of candidate fundamental theories, 
such as new symmetries, 
novel particles, 
or large extra dimensions, 
are also unlikely to require a low-energy description 
beyond effective field theory 
and can therefore be implemented into the SME, 
if necessary \cite{bek}. 
The above discussion indicates 
that dispersion-relation phenomenology 
offers at best narrow advantages in generality 
when compared to the SME. 

{\bf Advantages of the SME.}
The SME 
allows the identification 
and direct comparison 
of virtually all presently feasible Lorentz and CPT tests. 
Moreover, 
the SME contains 
classical kinematics test models of relativity 
(such as Robertson's framework, 
its Mansouri-Sexl extension, 
or the $c^2$ model)
as limiting cases \cite{km02}. 
Another advantage of the SME 
concerns the implementation 
of additional desirable conditions 
besides observer Lorentz invariance. 
For instance, 
one can choose to impose 
translation invariance, 
power-counting renormalizability, 
SU(3)$\times$SU(2)$\times$U(1) gauge symmetry, 
hermiticity,
and point-like interactions, 
which further restricts the parameter space for Lorentz breaking. 
As in the dispersion-relation case, 
one can adopt simplifying choices, 
such as rotational invariance 
in certain frames. 
This latter simplification of the SME 
has been assumed in Ref.\ \cite{cg99}. 

Concerning threshold investigations, 
the quadratic, translationally invariant sector 
of the SME 
determines possible one-particle dispersion relations, 
so that the correction \rf{ansatz} is constrained. 
Further restrictions are obtained 
by implementing the general conditions 
discussed in the previous paragraph. 

For example, 
the free renormalizable gauge-invariant contribution 
to the photon sector of $\de {\cal L}$ is given by \cite{ck98}
\beq
\de {\cal L}\;\supset\; (k_{AF})_{\mu}A_{\nu}\tilde{F}^{\mu\nu}
-\fr{1}{4}(k_{F})_{\mu\nu\rh\si}F^{\mu\nu}F^{\rh\si}\; ,
\label{sampleterm}
\eeq
where $A_{\nu}$ is the usual four-potential, 
$F^{\mu\nu}=\prt^{\mu}A^{\nu}-\prt^{\nu}A^{\mu}$ 
and $\tilde{F}^{\mu\nu}=\half\ve^{\mu\nu\rh\si}F_{\rh\si}$ 
denote the field-strength tensor and its dual, respectively,
and $(k_{AF})_{\mu}$ and $(k_{F})_{\mu\nu\rh\si}$ 
are constant parameters 
controlling the Lorentz violation. 
If in addition rotation invariance is assumed for the photon, 
suitable transformations 
move the Lorentz violation due to the $(k_{F})$ term 
into other sectors of the theory 
and map $(k_{AF})^{0}\rightarrow(\tilde{k}_{AF})^{0}$ \cite{km02}, 
so that only one Lorentz-violating parameter 
$\xi=(\tilde{k}_{AF})^{0}$ remains. 
Then, the usual methods yield
\beq
(\la^{\mu}\la_{\mu})^2+4\xi^2\la^{\mu}\la_{\mu}-4(\xi\la_0)^2=0\; .
\label{phdr}
\eeq
Equation \rf{phdr} 
is only one example of many other natural dispersion relations 
never considered 
in UHECR analyses. 

As an additional advantage, 
the SME provides the basis for the calculation of reaction rates, 
a determining factor for observational relevance. 
An example is given 
by vacuum \v{C}erenkov radiation \cite{cg99,cer}. 
The above discussion supports Result (iv):
{\it Threshold analyses within the SME 
ensure compatibility with dynamics and 
permit a much broader scope, 
while maintaining greatest possible generality}
\cite{fn8}. 
The above result strongly suggests 
that particle-reaction investigations 
are best performed in the context of the SME. 

\section{Positivity and causality}

Positivity and causality are fundamental principles 
in conventional physics 
and have been investigated in the Lorentz-violating context \cite{kl01}. 
However, we are not aware of any threshold analysis 
discussing these principles carefully. 
Therefore it seems appropriate 
to review some important aspects of positivity and causality 
before investigating 
the implementation of these properties 
into particle-reaction studies.

A postulate of Special Relativity states 
that velocities 
of material bodies and radiation 
are limited by the vacuum light speed. 
This postulate both contains a notion of causality 
and leads to the Lorentz transformations. 
This fact results in the common misconception 
that Lorentz breaking 
implies superluminal propagation, 
and hence causality violations. 
However, in many conventional situations 
involving a background, 
Lorentz symmetry is broken 
but causality is maintained. 
For instance, 
the anisotropic propagation of electromagnetic waves inside certain crystals 
is normally causal \cite{die,ck98}.  
Moreover, in such a situation the total conserved energy 
is clearly positive definite for all observers. 
Thus, 
the principles of positivity and causality 
are {\it a priori} independent and distinct 
from the principle of Lorentz symmetry. 
On the one hand, 
positivity and causality 
lead, e.g., to the spin-statistics theorem \cite{pau40,pct},  
which is a cornerstone of quantum field theory. 
On the other hand, 
polynomial Lorentz-violating dispersion relations 
fail to satisfy these requirements
above scales associated with the underlying theory \cite{kl01}. 
It is therefore natural to ask 
whether positivity or causality violations 
become acceptable in the presence of Lorentz breaking. 

{\bf Positivity.}
Lorentz violation 
introduces positivity problems  
through spacelike four-momenta for particles \cite{kl01}, 
which can then have negative energies in an arbitrary frame. 
In general, 
these negative energies 
can neither be eliminated by a shift 
of the energy zero 
nor by a coordinate-independent reinterpretation 
as corresponding to antiparticles. 
This loss of positivity 
can lead to 
unconventional instabilities
for one-particle states \cite{kl01}, 
but it may leave unaffected the stability of the vacuum. 
Note also the existence of a mechanism 
that avoids positivity problems 
despite the presence of spacelike momenta \cite{klp03}, 
so that such particles could potentially become admissible 
from a purely phenomenological viewpoint.

In light of the previous section, 
it is also necessary 
to investigate 
whether spacelike states 
can be incorporated into quantum field theory. 
Although it may be feasible 
to construct a Fock space of spacelike particles \cite{fein}, 
we are unaware 
of any internally consistent interacting quantum field theories 
involving such particles as asymptotic states. 
The usual assumptions in perturbation theory, 
for example, 
seem to exclude negative energies and spacelike on-shell momenta. 

{\bf Causality.}
Causality breakdown in the present context occurs 
for particles with superluminal group velocities \cite{kl01}. 
For such particles, 
the chronology of events, 
and in particular cause and effect, 
become observer-dependent concepts. 
Note also that in most frames, 
particle propagation forward and backward in time 
can become possible. 
It then appears difficult 
to make meaningful physical predictions. 
We mention, however, 
that the causality breakdown 
due to small Lorentz-violating effects 
appears mild enough 
to permit the existence of concordant frames, 
in which particle propagation is always forward in time, 
so that closed causal loops
and therefore many causal paradoxes 
could be avoided. 

Although superluminal particles 
may potentially be acceptable for concordant observers, 
it is unlikely 
that this type of causality breakdown 
can be accommodated within the framework 
of relativistic quantum field theory. 
Generally, 
a hermitian hamiltonian for massive fermions 
fails to exist in the majority 
of frames \cite{kl01}. 
In addition, 
the usual covariant perturbative expansion 
relies on time ordering, 
an operation no longer coordinate invariant, 
when microcausality is violated \cite{green02}. 
Again, it appears that privileged observers 
would have to be introduced 
to apply the conventional methods. 

{\bf Implications for threshold analyses.}
From a conservative viewpoint, 
it is natural to ask 
whether reaction-threshold kinematics is significantly affected 
if positivity and causality are imposed. 
Let $M$ and $m$ denote the scales 
of the fundamental theory and present-day low-energy physics, 
respectively. 
Then, 
the scale $p_{s{\textrm -}c}$ for the occurrence of spacelike momenta 
or causality problems 
in the case for dimension four operators 
can be as low as \cite{kl01}
\beq
p_{s{\textrm -}c}\sim{\cal O}(\sqrt{mM}\:)\; .
\label{scale}
\eeq
For instance, 
if $m$ is the proton mass 
and $M$ is taken to be the Planck scale, 
then $p_{s{\textrm -}c}\sim3\times 10^{18}$ eV. 
UHECRs with a spectrum extending beyond $10^{20}$ eV 
are often employed to bound Lorentz violation 
or to suggest evidence for Lorentz breaking. 
Thus, imposing positivity and causality 
could require modifications in threshold analyses. 

As specific example, 
consider the process  
$\ga\rightarrow \pi^0+\ga$ 
kinematically forbidden in conventional physics. 
The usual  dispersion-relation modifications 
in the literature permitting this process 
are associated with positivity or causality violations: 
Take the photon energy $E$ to be  
$E=({\vec{p}}^{\: 2}+\de f(\vec{p}))^{1/2}$.  
Here,  $\de f(\vec{p})$ 
excludes a mass term \cite{fn1} 
and depends only on the photon three-momentum $\vec{p}$. 
Then the point $(E,\vec{p})=(0,\vec{0})$ on the momentum-space lightcone 
satisfies the modified dispersion relation. 
For some three-momenta $\vec{p}\neq\vec{0}$, 
a nontrivial $\de f(\vec{p})$ forces $E(\vec{p})$ 
to curve to the outside or inside of the lightcone 
leading to spacelike four-momenta or superluminal group velocities, 
respectively. 
But taking the photon dispersion relation to be conventional 
requires spacelike pions for the decay to occur \cite{fn2}. 

The above arguments lead to Result (v): 
{\it Positivity and causality should be maintained 
in threshold investigations.} 
To enforce positivity and causality 
one could perform all reaction analyses 
well below the breakdown scale for these properties. 
However, 
in many practical applications 
this approach may be unsatisfactory 
because of the limited accessible momentum range. 
It thus seems desirable 
to employ dispersion relations for Lorentz violation 
that preserve positivity and causality at all scales. 
Such dispersion relations 
can be motivated in string field theory 
and have been investigated in the context of the SME \cite{kl01}.

\section{Further results} 

In this section, 
we consider additional, important kinematics issues 
that are best discussed by example. 
We focus on the rotationally invariant $n=3$ case, 
which has received a lot of attention in the literature 
\cite{alo,gac,jlm,koma,alf,posp}. 
We also take this opportunity 
to illustrate some of our results from the previous sections. 

{\bf The $|\vec{\la}|^3$ correction.} 
The standard approach to $n=3$ dispersion relations 
assumes 
\beq
{\la_0}^2-\vec{\la}^{\:2}=m^2+\fr{|\vec{\la}|^{\,3}}{M}\; ,
\label{p3dr}
\eeq
Note that Result (iii) implies 
that \rf{p3dr} has problems with coordinate invariance. 
Note also the fourfold degeneracy 
of the energies, 
an undesirable feature in light of Result (ii). 
Moreover, 
it will become clear below 
that dispersion relations of the type \rf{p3dr} 
develop positivity or causality violations 
at energies 
that are phenomenologically accessible, 
when $M$ is taken to be a Planck-scale mass. 

We first consider a positive $M$. 
Equation \rf{p3dr} has the roots 
${\la_0}_{\pm}=\pm(|\vec{\la}|^{\,3}/M+\vec{\la}^{\:2}+m^2)^{1/2}$. 
We proceed by assuming 
that the usual reinterpretation 
of the negative-energy roots ${\la_0}_-$ is still applicable, 
so that these solutions 
correspond to positive-energy reversed-momentum particle states: 
\bea
p^{\mu}_+ = (E_+,\vec{p})\; &,&
\quad E_+(\vec{p})={\la_0}_{+}(\vec{p})\; ,\nonumber\\
p^{\mu}_- = (E_-,\vec{p})\; &,&
\quad E_-(\vec{p})=-{\la_0}_{-}(-\vec{p})\; .
\label{reint}
\eea
The respective particle and antiparticle energies $E_+$ and $E_-$ 
are then given by 
\beq 
E_{\pm}(\vec{p})=\sqrt{\fr{|\vec{p}|^{\,3}}{M}+\vec{p}^{\:2}+m^2}\; . 
\label{eigenen} 
\eeq 
Note that spacelike particle momenta are absent. 
The corresponding group velocities 
$\vec{v}^{\;\pm}_g=\prt E_{\pm}/\prt \vec{p}$ 
obey 
\beq
\vec{v}^{\;\pm}_g(\vec{p})=
\fr{1+3|\vec{p}|/2M}{\sqrt{{|\vec{p}|^{\,3}}/{M}+\vec{p}^{\:2}+m^2}}
\;\vec{p}\; .
\label{gv1}
\eeq
It can be shown  
that above momentum scales $p_c\simeq\sqrt[3\ ]{m^2M/2}$ 
both the particle and the antiparticle 
develop superluminal group velocities violating causality. 

Suppose we had chosen the correction term ${|\vec{\la}|^{\,3}}/{M}$ 
to enter the dispersion relation \rf{p3dr} 
with a minus instead of a plus sign. 
The resulting particle energies and group velocities 
are then obtained by the replacement $M\rightarrow -M$ 
in Eqs.\ \rf{eigenen} and \rf{gv1}, respectively. 
In this case, 
superluminal group velocities are absent, 
but positivity problems due to spacelike momenta 
occur at scales $p_s=\sqrt[3\ ]{m^2M}$. 
At three-momenta $|\vec{p}|>2M/3$ 
the particle energy decreases with increasing $|\vec{p}|$, 
so that monotonicity of $E_{\pm}(|\vec{p}|)$ is lost. 
Moreover, at momenta above the scale $M$, 
dispersion relation \rf{p3dr} admits imaginary particle energies.  
Both of these features 
signal the breakdown of validity of \rf{p3dr}. 
However, if $M$ is of the order of the Planck mass, 
the required particle momenta are phenomenologically uninteresting. 

In the context of \rf{p3dr}, 
we next consider the photon decay $\ga\rightarrow e^++e^-$ 
into an electron-positron pair, 
which is kinematically forbidden in conventional physics. 
For all particles, 
we assume the same $M$ parameter, 
and we set $m=0$ for the photon. 
Energy-momentum conservation requires 
\beq
k^{\mu}=p^{\mu}_++p^{\mu}_-\; ,
\label{emc}
\eeq
where $k^{\mu}$ is the photon four-momentum,
and $p^{\mu}_+$ and $p^{\mu}_-$ 
are the four-momenta of the positron and electron, 
respectively. 
In App.\ A, 
we show rigorously 
that the decay is kinematically allowed 
provided the photon three-momentum obeys $|\vec{k}|\ge\sqrt[3\ ]{4m^2M}$. 
Then, causality problems occur 
because at least one decay product must be superluminal. 
If a dispersion relation of the type \rf{p3dr} 
but with the opposite sign of the correction term is used, 
the above photon-decay process is kinematically forbidden 
within the validity range $|\vec{p}|<2M/3$. 
This fact is also demonstrated rigorously in  App.\ A. 

{\bf The $\pm|\vec{\la}|^3$ correction.} 
Consider now enforcing observer Lorentz invariance 
by allowing two simultaneous signs for the correction term in \rf{p3dr}. 
The particle and antiparticle energies are then 
\beq 
E_{\pm}^{(\al)}(\vec{p})=\sqrt{(-1)^{\al}
\fr{|\vec{p}|^{\,3}}{M}+\vec{p}^{\:2}+m^2}\; , 
\label{eigenen3} 
\eeq 
where $\al=1,2$ labels the two possible particle (antiparticle) energies, 
which perhaps correspond to different spin-type states. 
As a result, 
six kinematically distinct decays have to be considered. 
Note, however, 
that additional conservation laws associated with rotational invariance 
may preclude some of the six reactions. 
A proper investigation of this case 
therefore requires dynamical concepts. 
This is our Result (vi): 
{\it The effects of assumed symmetries, 
such as rotational invariance, 
must be incorporated into threshold analyses.}

{\bf The $\la_0\vec{\la}^{\:2}$ correction.}
Another $n=3$ coordinate-independent dispersion relation is 
\beq
{\la_0}^2-\vec{\la}^{\:2}=m^2+\fr{\la_0\vec{\la}^{\:2}}{M}\; .
\label{foamdr}
\eeq
The usual reinterpretation \rf{reint}
yields the respective 
particle and antiparticle energies 
$E_+$ and $E_-$: 
\beq 
E_{\pm}(\vec{p})=\sqrt{\fr{\vec{p}^{\:4}}{4M^2}+\vec{p}^{\:2}+m^2} 
\; \pm \fr{\vec{p}^{\:2}}{2M}\; . 
\label{eigenen2} 
\eeq 
Note that the particle-antiparticle degeneracy is lifted, 
and only the antiparticles develop spacelike momenta, 
and thus difficulty with positivity, 
above the scale $p_s=\sqrt[3\ ]{m^2M}$. 
The particle momenta $(E_+,\vec{p})$ 
remain timelike for all $\vec{p}\neq \vec{0}$. 
The corresponding group velocities are  
\beq
\vec{v}^{\;\pm}_g(\vec{p})=\left(\fr{\vec{p}^{\:2}+2M^2}{\sqrt{
{\vec{p}^{\:4}}+{4M^2}\vec{p}^{\:2}+{4M^2}m^2}}\;
\pm 1\right)\fr{\vec{p}}{M}\; .
\label{gv}
\eeq 
Thus, 
at any given nonzero three-momentum 
the particle travels faster than the antiparticle. 
Moreover, 
one can verify 
that above the scale $p_c\simeq\sqrt[3\ ]{m^2M/2}$ 
the particle speed becomes superluminal 
leading to causality problems. 
The antiparticle always remains subluminal. 
For $|\vec{p}|\gg M$, 
the antiparticle's speed goes to zero, 
a feature indicating the validity breakdown 
of \rf{foamdr}. 
In any case, 
this momentum range 
appears phenomenologically uninteresting at the present time. 

Consider again photon decay into an electron-positron pair. 
We now assume a dispersion relation of the type \rf{foamdr}, 
take the lepton and photon corrections 
to be controlled by the same parameter $M$, 
and set $m=0$ for the photon. 
For notational simplicity 
we define the positron as the particle 
and the electron as the antiparticle. 
Because of the two possible incoming photon states $\ga_+$ and $\ga_-$, 
two kinematically distinct processes must be investigated. 
The subscripts $+$ and $-$ 
correspond to ones for the particle energy in \rf{eigenen2}. 
In App.\ B, 
it is demonstrated 
that the decay $\ga_{+}\rightarrow e^++e^-$ is allowed 
above a certain threshold. 
If the observed value $m=0.511$ MeV for the electron and positron masses 
is used, 
and $M$ is taken to be the Planck mass, 
the numerically determined threshold value 
for the incoming photon three-momentum 
is $|\vec{k}_{\rm min}|\simeq 7.21$ TeV. 
Appendix B also contains a proof 
that the decay channel in question 
is kinematically forbidden for $\ga_-$ photons. 

Suppose we had chosen the Lorentz-violating correction 
to enter the dispersion relation 
with the opposite sign. 
In such a situation, 
the roles of particle and antiparticle 
are interchanged. 
However, 
apart from this trivial reinterpretation, 
the above discussion remains unaffected. 
In particular, 
the threshold for photon decay into an electron-positron pair 
is left unchanged. 

The above example demonstrates 
that, 
contrary to claims in the literature, 
a $\la_0\vec{\la}^{\:2}$ correction 
kinematically permits photon decay 
irrespective of the sign of the correction. 
In addition, 
the discussion in this section 
identifies the common assumption 
$\la_0\simeq|\vec{\la}|$ 
as the source of this confusion. 
In a general context, 
we arrive at Result (vii): 
{\it In threshold analyses, 
many approximations, 
such as $\la_0\simeq|\vec{\la}|$, 
and others leading to additional degeneracies 
are typically invalid.} 
We remark that Result (vii) 
applies also to the conventional case. 

\section{Conclusion}

This work has considered 
Lorentz-violating dispersion-relation modifications 
and some of their implications 
in the search for possible signatures 
for fundamental physics. 
More specifically, 
we have discussed 
the role of a dynamical framework 
and the requirements 
of coordinate independence, 
positivity, 
and causality 
in the subject. 

The consequences of these principles  
are summarized in the various Results (i) to (vii) 
given in the text. 
Correct threshold investigations within the SME 
are automatically compatible with these requirements. 

None of the particle-reaction analyses known to the author 
is consistent with all of the Results (i) to (vii). 
It would therefore be of great interest 
to revisit many threshold studies 
implementing the findings of the present work.

\section*{Acknowledgments} 

This work was supported in part 
by the Centro Multidisciplinar de Astrof\'{\i}sica (CENTRA).

\begin{appendix}

\section{Thresholds in the $|\vec{\la}|^3$ case}

Implementing momentum conservation, 
we write $\vec{k}$, $\vec{p}$, and $\vec{k}-\vec{p}$ 
for the photon, positron, and electron three-momenta, 
respectively. 
Then, 
the dispersion relation \rf{p3dr} 
(for the photon with $m=0$) 
yields the respective particle energies 
$k_0(\vec{k})$, $p_0(\vec{p})$, and $p_0(\vec{k}-\vec{p})$, 
which obey the conservation equation 
\beq
k_0(\vec{k})=p_0(\vec{p})+p_0(\vec{k}-\vec{p})\; .
\label{encons} 
\eeq
At $\vec{k}=\vec{0}$, 
the reaction is forbidden 
because the left-hand side of \rf{encons} 
is smaller than the right-hand side for all $\vec{p}$. 
To find the threshold for the reaction, 
one can imagine to increase $|\vec{k}|$ 
until $k_0(\vec{k})$ becomes large enough 
to equal the minimum of the right-hand side of \rf{encons} 
viewed as a function $\vec{p}$. 
The energy of the decay products is smallest, 
when the outgoing three-momenta are parallel \cite{mjl}: 
Suppose adding transverse momenta 
to $\vec{p}$ and $\vec{k}-\vec{p}$ 
taken as parallel. 
This would increase the total outgoing energy 
because $p_0$ is strictly monotonic for positive arguments. 
A similar reasoning can be employed 
to exclude antiparallel final configurations. 

Next, we show that for a given $\vec{k}$, 
the minimum of the outgoing energy is attained, 
when the three-momentum is shared equally 
by the decay products, 
i.e., $\vec{p}=\vec{k}/2$. 
Equation \rf{threseq} would then 
yield the threshold quoted in the text. 
Using the expression for the particle energies \rf{eigenen} 
and implementing the above considerations 
yields: 
\beq
\sqrt{\fr{k^{3}}{M}+k^2}=\sqrt{\fr{p^{3}}{M}+p^2+m^2}
+(\:p\leftrightarrow p-k\:)\; ,
\label{threseq}
\eeq
where $k=|\vec{k}|$ and $p=|\vec{p}|\le k$.
The first derivative of the right-hand side of \rf{threseq} 
with respect to $p$ vanishes at $p=k/2$ 
consistent with the presence of an extremum. 
To complete the argument, 
we have to confirm  
that $p=k/2$ is the location of a minimum 
and that for $p\in[0,k]$ 
all values of the right-hand side of \rf{threseq} 
lie above the one at $p=k/2$. 
This will be the case 
if the second derivative of right-hand side of \rf{threseq} 
with respect to $p$ is nonnegative on $[0,k]$. 

Consider the positron energy 
and take the second derivative: 
\beq
\fr{\prt^2}{\prt p^2}\;p_0(p)=g(p)-h(p)\; .
\label{2der}
\eeq
Here, the functions $g$ and $h$ are given by
\bea
g(p)&=&\fr{3p+M}{(Mp^3+M^2p^2+M^2m^2)^{1/2}}\nonumber\\
h(p)&=&\fr{M(3p^2/2+Mp)^2}{(Mp^3+M^2p^2+M^2m^2)^{3/2}}\; .
\label{gh}
\eea
To see that $g>h$ for $p\in[0,k]$, 
we begin with the trivial inequality
$0<3p^4+4Mp^3+12Mm^2p+4M^2m^2$. 
Addition of $(3p^2+2Mp)^2$ to both sides of this inequality yields  
$(3p^2+2Mp)^2<4(3p+M)(p^3+Mp^2+Mm^2)$. 
We finally multiply 
both sides with $M/4(Mp^3+M^2p^2+M^2m^2)^{3/2}$ 
to obtain $\prt^2 p_0(p)/\prt p^2>0$. 
Replacing $p\rightarrow k-p$ in the above argument 
shows that $\prt^2 p_0(k-p)/\prt p^2>0$. 
This establishes the positivity 
of the second derivative of the right-hand side of \rf{threseq} 
with respect to $p$, 
which concludes our proof. 

We next consider the case with the opposite sign in \rf{p3dr} 
and show 
that the decay $\ga\rightarrow e_++e_-$ is kinematically forbidden 
within the validity range $0<k,p<2M/3$. 
Since monotonicity still holds, 
similar considerations as above imply 
that we must demonstrate 
\beq
\!\!\sqrt{-\fr{k^{3}}{M}+k^2}<\sqrt{-\fr{p^{3}}{M}+p^2+m^2}
+(\:p\leftrightarrow p-k\:)
\label{threseq2}
\eeq
for any value of $k$ and $p$ 
less than $2M/3$. 
Let us denote the right-hand side of the above inequality \rf{threseq2} 
by $R(k,p)$. 
We observe 
that in the validity range, 
$R$ can be decreased setting $m=0$: 
\beq
\sqrt{p^2-\fr{p^{3}}{M}}+\sqrt{(k-p)^2-\fr{(k-p)^{3}}{M}}<R(k,p)\; .
\label{R}
\eeq
Inequality \rf{R} will continue to hold, 
when its left-hand side $L(k,p)$ is minimized with respect to $p$. 
If we can demonstrate 
that the minimum is attained at either $p=0$ or $p=k$, 
then \rf{R} reduces to \rf{threseq2}, 
and we are done. 

We will show 
that $L(k,p)$ has a local maximum at $p/2$ 
and decreases monotonically in directions away from the maximum. 
Indeed, the derivative of $L$ with respect to $p$ 
vanishes at $p/2$. 
Negativity of the second derivative of $L$ 
would establish the desired result. 
One can show that  
\beq
\fr{\prt^2}{\prt p^2}\;L(k,p)=G(p)-H(p)+(p\:\leftrightarrow\; k-p)\; ,
\label{2der2}
\eeq
where the functions $G$ and $H$ are defined by 
\bea
G(p)&=&\fr{M-3p}{(M^2p^2-Mp^3)^{1/2}}\; ,\nonumber\\
H(p)&=&\fr{M(Mp-3p^2/2)^2}{(M^2p^2-Mp^3)^{3/2}}\; . 
\label{gh2}
\eea
We start from $0>-p^3(4M-3p)$, 
which holds within the validity range. 
Adding $(2Mp-3p^2)^2$ to both sides of this inequality 
yields $(2Mp-3p^2)^2<4(Mp^2-p^3)(M-3p)$. 
Multiplication with $M/(M^2p^2-Mp^3)^{3/2}$ 
shows that $G(p)<H(p)$. 
Since $k-p$ is also within the range of validity, 
we can infer $G(k-p)<H(k-p)$. 
Together with \rf{2der2}, 
these results establish the negativity claim. 

\section{Thresholds in the $\la_0\vec{\la}^{\:2}$ case}

To see that the decay is permitted, 
consider the special case of the decay $\ga_+\rightarrow e^++e^-$, 
in which the $e^+$ particle has zero three-momentum 
and the $\ga_+$ and $e^-$ particles both have three-momentum $\vec{k}$. 
Energy conservation and Eq.\ \rf{eigenen2} then imply 
\beq
\sqrt{\fr{k^4}{4M^2}+k^2}
=m+
\sqrt{\fr{k^4}{4M^2}+k^2+m^2}\;-\fr{k^2}{M}\; ,
\label{encons3}
\eeq
where $k=|\vec{k}|$, as before. 
Note that at $k=0$, 
the left-hand of \rf{encons3} 
is smaller than the right-hand side. 
For large $k$, 
the situation is vice-versa: 
The left-hand side grows quadratically, 
whereas the right-hand side decreases quadratically. 
By continuity, 
there must be some value of $k$, 
for which \rf{encons3} is satisfied, 
and the decay is allowed. 

For completeness, 
we demonstrate that the decay of $\ga_-$ 
is forbidden in the present context. 
Arguments similar to the ones in App.\ A imply 
that we have to show the validity of  
\bea
\sqrt{\fr{k^4}{4M^2}+k^2}
&<&\sqrt{\fr{p^4}{4M^2}+p^2+m^2}+\fr{(k-p)}{M}\:k
\nonumber\\
&&{}+\sqrt{\fr{(k-p)^4}{4M^2}+(k-p)^2+m^2}
\label{cond}
\eea
for $p\in[0,k]$, 
where $p=|\vec{p}|$ is the three-momentum magnitude 
of the $e^-$ particle. 
The right-hand side $r(k,p)$ of \rf{cond} 
can be reduced by replacing in the second term 
the factor of $k$ by $p$ 
and setting $m=0$ under the square roots: 
\beq
\!\!\sqrt{\fr{p^4}{4M}+p^2}+(\: p\leftrightarrow k-p\: )+\fr{(k-p)}{M}\:p
<r(k,p)\; .
\label{cond2}
\eeq
This inequality remains true, 
when the left-hand side 
is minimized with respect to $p$. 
If we can show 
that the minimum is attained at $p=0$ or at $p=k$, 
the inequalities \rf{cond2} and \rf{cond} become identical, 
and the claim follows. 

We will establish 
that $r(k,p)$ has a local maximum at $p=k/2$ 
and decreases in directions away from this maximum. 
The necessary condition 
$\prt r/\prt p=0$ at $p=k/2$ can be verified straightforwardly. 
We also show that $\prt^2 r/\prt p^2<0$ 
for $p\in[0,k]$. 
An explicit expression involving the second derivative is given by:
\beq
M\fr{\prt^2}{\prt p^2}r(k,p)=F(p)+F(k-p)-2\; ,
\label{rexp2}
\eeq
where the function $F$ is defined by 
\beq
F(p)=\fr{p(6M^2+p^2)}
{(p^2+4M^2)^{3/2}}\; .
\label{rexp3}
\eeq
Thus, 
it suffices to prove $F<1$. 
We observe that $0<64M^6p^2+12M^4p^4$ 
and add $p^4(6M^2+p^2)^2$ 
to both sides of this inequality. 
Then, the right-hand side 
can be cast into the form $p^2(p^2+4M^2)^3$. 
Dividing by this expression 
yields $F^2<1$, 
which completes the proof.

\end{appendix}

\end{multicols}

\begin{references}
\bibitem{comment} 
In the context of the renormalizable sector of the SME, 
dispersion relations are typically fourth-order polynomials 
in $\la_0$, 
so that \rf{lvdr} together with ansatz \rf{ansatz} is inconvenient. 
However, 
if $\de f$ is allowed to contain unsuppressed terms, 
SME dispersion relations can be generated.
\bibitem{fn3} 
Negative or fractional powers of $|\vec{\la}|$ could arise, 
when the dispersion relation \rf{lvdr} 
with the ansatz \rf{ansatz} is rearranged, 
for example, 
if it is solved for $\la_0=\la_0(\vec{\la})$.
\bibitem{fn6}
Odd powers of $|\vec{\la}|$ 
could arise from equations of motion 
containing nonlocal operators of the type $\sqrt{-\Delta}$, 
where $\Delta$ is the laplacian. 
However, such nonlocalities are presently unmotivated 
in candidate fundamental theories, 
so that we exclude such operators 
in the present work.
\bibitem{fn8}
For explicit photon and fermion dispersion relations, 
see Refs.\ \cite{ck98} and \cite{kl01}, respectively.
Fermion-eigenenergy approximations can be found in 
R.\ Lehnert,
UMI-30-54449.
\bibitem{fn1} 
A mass term would yield a Lorentz-invariant 
(but gauge-symmetry violating) contribution, 
which is not of interest in the present context.
\bibitem{fn2} 
The allowed phase space for the decay products 
in case of a lightlike pion four momentum 
is a set of measure zero 
leading at best to a suppressed rate for the reaction.
\end{references}
\end{document}